\documentclass[twocolumn,aps,prb,superscriptaddress,amsmath,amssymb,superscriptaddress]{revtex4-1}
\newcommand{\pagenumbaa}{1}
\usepackage{graphicx}
\usepackage{color}
\usepackage{braket}

\usepackage{verbatim}
\usepackage{amsmath}
\usepackage{mathtools}
\DeclarePairedDelimiter\abs{\lvert}{\rvert}
\usepackage{float}
\usepackage{hyperref}
\usepackage{lipsum}

\begin{document}
\title{Radiative Auger Process in the Single-Photon Limit}
\author{Matthias C. L\"obl}
\email[]{matthias.loebl@unibas.ch}
\affiliation{Department of Physics, University of Basel, Klingelbergstrasse 82, CH-4056 Basel, Switzerland}
\author{Clemens Spinnler}
\affiliation{Department of Physics, University of Basel, Klingelbergstrasse 82, CH-4056 Basel, Switzerland}
\author{Alisa Javadi}
\affiliation{Department of Physics, University of Basel, Klingelbergstrasse 82, CH-4056 Basel, Switzerland}
\author{Liang Zhai}
\affiliation{Department of Physics, University of Basel, Klingelbergstrasse 82, CH-4056 Basel, Switzerland}
\author{Giang N. Nguyen}
\affiliation{Lehrstuhl f\"ur Angewandte Festk\"orperphysik, Ruhr-Universit\"at Bochum, DE-44780 Bochum, Germany}
\affiliation{Department of Physics, University of Basel, Klingelbergstrasse 82, CH-4056 Basel, Switzerland}
\author{Julian Ritzmann}
\affiliation{Lehrstuhl f\"ur Angewandte Festk\"orperphysik, Ruhr-Universit\"at Bochum, DE-44780 Bochum, Germany}
\author{Leonardo Midolo}
\affiliation{\mbox{Niels Bohr Institute, University of Copenhagen, Blegdamsvej 17, Copenhagen DK-2100, Denmark}}
\author{Peter Lodahl}
\affiliation{\mbox{Niels Bohr Institute, University of Copenhagen, Blegdamsvej 17, Copenhagen DK-2100, Denmark}}
\author{Andreas D. Wieck}
\affiliation{Lehrstuhl f\"ur Angewandte Festk\"orperphysik, Ruhr-Universit\"at Bochum, DE-44780 Bochum, Germany}
\author{Arne Ludwig}
\affiliation{Lehrstuhl f\"ur Angewandte Festk\"orperphysik, Ruhr-Universit\"at Bochum, DE-44780 Bochum, Germany}
\author{Richard J. Warburton}
\affiliation{Department of Physics, University of Basel, Klingelbergstrasse 82, CH-4056 Basel, Switzerland}

\begin{abstract}
In a multi-electron atom, an excited electron can decay by emitting a photon. Typically, the leftover electrons are in their ground state. In a radiative Auger process, the leftover electrons are in an excited state and a red-shifted photon is created \cite{Aberg1969,Aberg1971,Bloch1935b,Bloch1935}. In a quantum dot, radiative Auger is predicted for charged excitons \cite{Michler2003}. Here, we report the observation of radiative Auger on trions in single quantum dots. For a trion, a photon is created on electron-hole recombination, leaving behind a single electron. The radiative Auger process promotes this additional (Auger) electron to a higher shell of the quantum dot. We show that the radiative Auger effect is a powerful probe of this single electron: the energy separations between the resonance fluorescence and the radiative Auger emission directly measure the single-particle splittings of the electronic states in the quantum dot with high precision. In semiconductors, these single-particle splittings are otherwise hard to access by optical means as particles are excited typically in pairs, as excitons. After the radiative Auger emission, the Auger carrier relaxes back to the lowest shell. Going beyond the original theoretical proposals, we show how applying quantum optics techniques to the radiative Auger photons gives access to the single-electron dynamics, notably relaxation and tunnelling. This is also hard to access by optical means: even for quasi-resonant $p$-shell excitation, electron relaxation takes place in the presence of a hole, complicating the relaxation dynamics. The radiative Auger effect can be exploited in other semiconductor nanostructures and quantum emitters in the solid state to determine the energy levels and the dynamics of a single carrier.
\end{abstract}

\maketitle

\setcounter{page}{\pagenumbaa}
\thispagestyle{plain}
Auger processes are a well-known phenomenon in atoms \cite{Bambynek1972,Barthes1995}. Nonradiative Auger processes involving continuum states have been observed in several solid-state systems: quantum dots \cite{Kurzmann2016}, two-dimensional materials \cite{Han2018}, colour centers \cite{Siyushev2013}, and semiconductor lasers \cite{Blood2015}. As originally predicted for atoms, an Auger process can also take place in connection with a radiative transition \cite{Bloch1935b,Bloch1935}. In such a radiative Auger process, part of the available energy is transferred to another electron and the emitted photon is correspondingly red-shifted. The radiative Auger process has been observed in X-ray spectra \cite{Aberg1969,Aberg1971}. The so-called electron shake-off process has a similar physical origin \cite{Carlson1963}. At optical frequencies, the radiative Auger process has been described in ensembles of donors \cite{Dean1967} and as a so-called shake-up process in the Fermi-sea \cite{Skolnick1994,Manfra1998,Kleemans2010}, a many-particle effect. On a single-photon emitter or in a few-electron configuration, the radiative Auger process has not been observed.

\begin{figure*}[t]
\includegraphics[width=1.8\columnwidth]{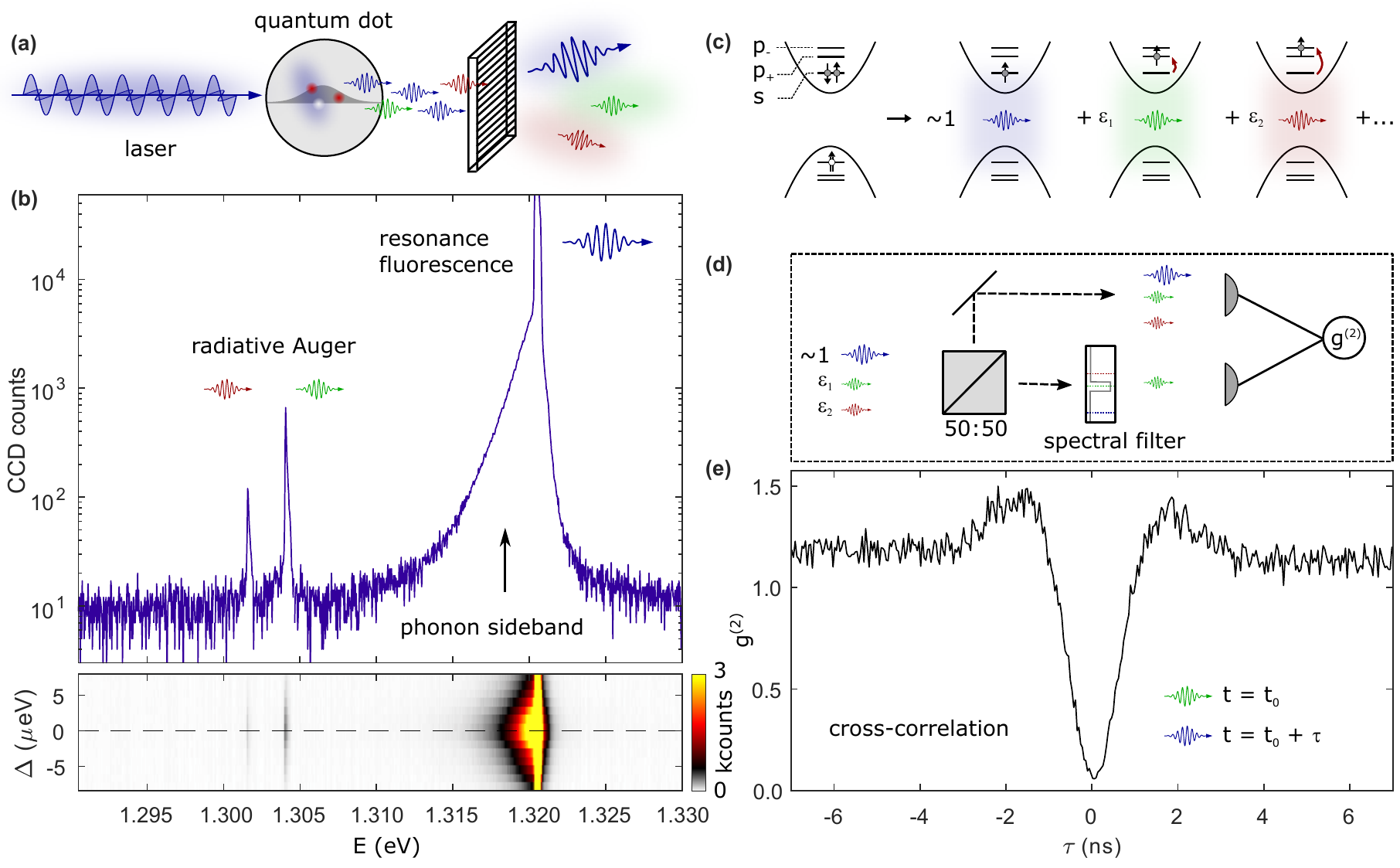}
\caption{\label{fig:mechanism}\textbf{Observation of a radiative Auger process on a single quantum dot. (a)} Schematic of the experimental setup: the quantum dot (QD) is resonantly excited with a narrow-bandwidth laser, and its emission is spectrally resolved. \textbf{(b)} Upper panel, emission spectrum of the negative trion ($X^{1-}$) in an InGaAs QD under resonant excitation ($T=4.2\ \text{K}$). The strong peak at $E\simeq1.321\ \text{eV}$ is the resonance fluorescence, which is surrounded by a broad LA-phonon sideband. Red-shifted by $\hbar\omega_0\sim18\ \text{meV}$ there are two additional emission lines, stemming from the radiative Auger process. Lower panel, the QD can be tuned in and out of the resonance with the laser by exploiting the dc Stark effect via a gate voltage, $V_g$. The shown spectrum is measured at zero detuning, $\Delta$, between QD and laser (dashed line). Resonance fluorescence and radiative Auger are maximum when QD and laser are in resonance ($\Delta=0$). (See also section \ref{sec:power}) \textbf{(c)} Mechanism of the radiative Auger process: with a probability close to one, the trion recombination results in an emission of a resonant photon and leaves the remaining electron in the ground state ($s$-shell). With small probabilities $|\epsilon_{1}|^2$ and $|\epsilon_{2}|^2$, the remaining electron is promoted into one of the $p$-shells, and the photon is consequently red-shifted. \textbf{(d)} Setup for the cross-correlation between the radiative Auger emission and the resonance fluorescence. The delay $\tau$ corresponds to the duration between the arrival of a resonant photon on detector 2 after the detection of an Auger photon on detector 1. \textbf{(e)} Cross-correlation measurement between the radiative Auger emission and the resonance fluorescence. The strong anti-bunching at zero time-delay proves that both emission lines originate from the same emitter.}
\end{figure*}

We observe the radiative Auger process on two different systems: first, a self-assembled InGaAs quantum dot (QD) in GaAs grown in the Stranski-Krastanov mode \cite{Michler2003} and second, a GaAs QD in AlGaAs grown by infilling of droplet-etched nano-holes \cite{Huo2013}. We resonantly excite the negative trion ($X^{1-}$) of a QD with a narrow-bandwidth laser. In both QD systems, the charge state of the QD is precisely controlled via Coulomb blockade \cite{Warburton2000}. We collect the emission of the QD and resolve it spectrally, as schematically shown in Fig.\ \ref{fig:mechanism}(a). Shown in Fig.\ \ref{fig:mechanism}(b) is the result of such a measurement for an InGaAs QD. The main peak at photon energy $\sim1.321\ \text{eV}$ is the resonance fluorescence of the trion. This spectrally narrow emission is accompanied by an LA-phonon sideband on the red side \cite{Hansom2014b,Koong2019,Brash2019}. In addition, we observe two weak emission lines, red-shifted by $\sim18\ \text{meV}$ from the main fluorescence peak. In the following, we show that these emission lines originate from a radiative Auger process as illustrated in Fig.\ \ref{fig:mechanism}(c): an electron and a hole recombine optically and with a small probability, the second electron is promoted to an excited state, the $p$-shell of the QD. In the case of resonance fluorescence, in contrast, the optical recombination of the trion leaves behind a single electron in the ground state ($s$-shell of the QD). 

\begin{figure*}[t]
\includegraphics[width=1.8\columnwidth]{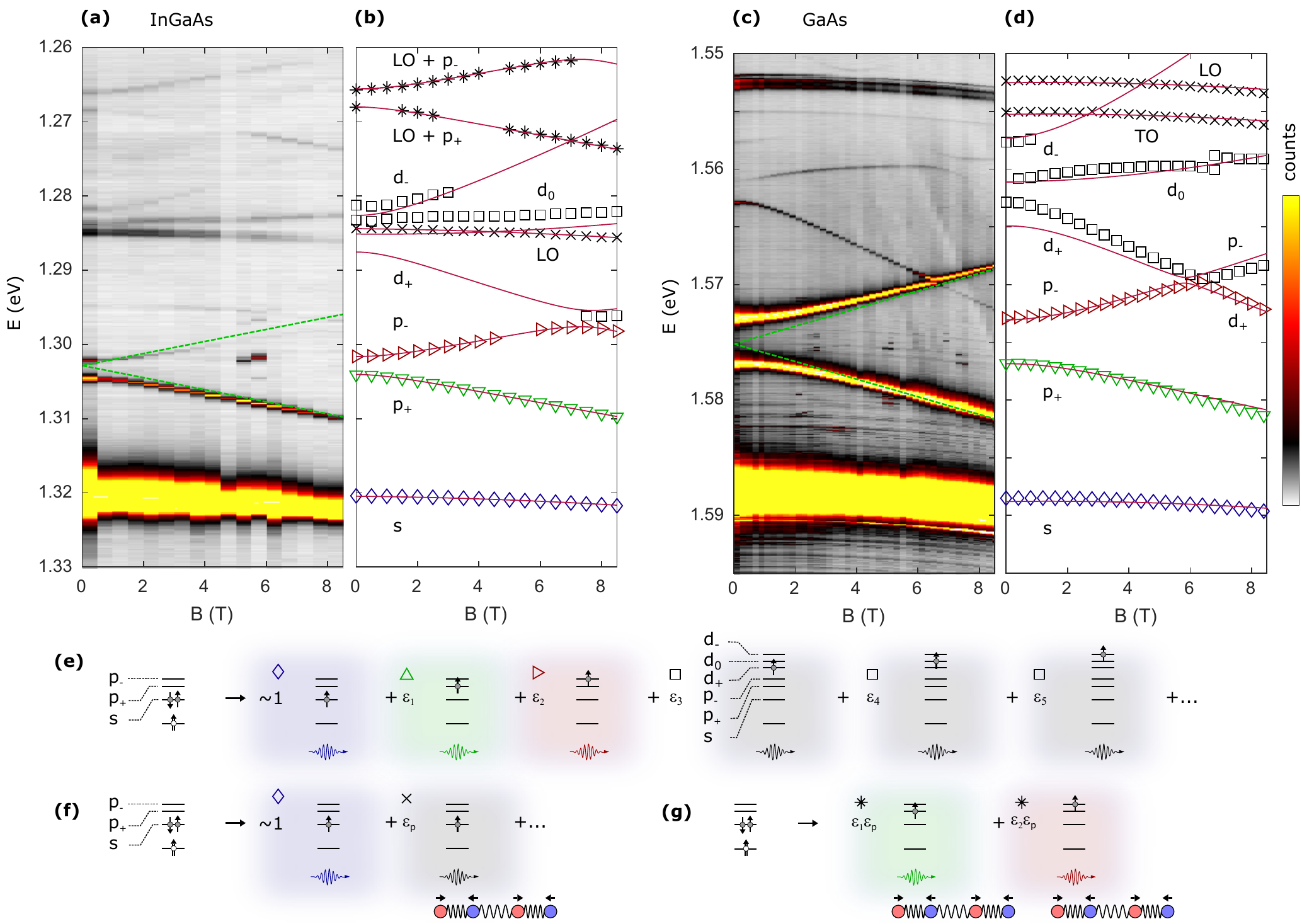}
\caption{\label{fig:dispersion}{\bf Magnetic field dispersion of the radiative Auger emission. (a)} Emission spectrum under resonant excitation as a function of the magnetic field measured on an InGaAs quantum dot (QD). The two green lines indicate the radiative Auger emission where one electron is promoted into the $p$-shells. This emission follows a dispersion of $\sim\pm\frac{1}{2}\hbar\omega_c$, with $m^*\simeq0.071\ m_e$ ({\color{black}$s$-to-$p$-splitting:} $\hbar\omega_0\simeq17.7\ \text{meV}$, further parameters in Tab. \ref{tab:fit}). {\bf(b)} Magnetic field dispersion of the radiative Auger emission. The emission lines above the $s$-shell can be well described by the Fock-Darwin spectrum. The red lines represent a fit of our analytical model of the radiative Auger emission. {\bf(c)} Radiative Auger emission as a function of the magnetic field measured on a GaAs QD ($m^*\simeq0.076\ m_e$, $\hbar\omega_0\simeq13.8\ \text{meV}$). {\bf(d)} Magnetic field dispersion of the radiative Auger emission for the GaAs QD. {\bf(e)} Schematics of the radiative Auger process involving both $p$- and $d$-shells. {\bf(f)} Optical recombination involving the creation of an LO- or a TO-phonon. We note that this process is observed for the trion and the neutral exciton (see Fig. \ref{fig:mechanismFull}). $|\epsilon_{p}|^2$ labels the probability for the process involving the LO phonon. {\bf(g)} Schematics of the radiative Auger process involving both carrier excitation to the $p$-shell and the creation of a phonon.}
\end{figure*}

Several observations substantiate the interpretation that the two red-shifted lines originate from a radiative Auger process. First, the Auger lines disappear on removing the additional electron -- they are absent in the emission spectrum of the neutral exciton, $X^0$ (see Fig. \ref{fig:mechanismFull}). Second, the red-shifted emission lines only appear when the laser is in resonance with the QD (Fig.\ \ref{fig:mechanism}(b)). Third, the time-resolved cross-correlation between the radiative Auger emission and the resonance fluorescence (Fig.\ \ref{fig:mechanism}(d,e)) shows a pronounced anti-bunching at zero time-delay. This measurement demonstrates that the different emission lines originate from the same QD. The emitter produces either a resonance-fluorescence photon or a radiative-Auger photon, but never two photons at the same time. Finally, to prove that the radiative Auger process leaves an electron in a higher shell, we measure the optical emission as a function of the magnetic field (Faraday geometry). The magnetic field dispersion of the radiative Auger emission is shown in Fig.\ \ref{fig:dispersion}(a,b) for an InGaAs QD and in Fig.\ \ref{fig:dispersion}(c,d) for a GaAs QD. At high magnetic fields, the two red-shifted emission lines, which are the closest in energy to the resonance-fluorescence, have a dispersion of $\pm\frac{1}{2}\hbar\omega_c$ (cyclotron frequency: $\omega_c=\frac{eB}{m^*}$, electron effective mass $m^*$). This magnetic field dispersion shows that the emission is connected to an energy transfer to the $p$-shells. More generally, the strong magnetic field dispersion of the radiative Auger emission arises because the magnetic field creates an additional orbital confinement, which leads to a strong magnetic field dependence of higher QD-shells \cite{Kouwenhoven2001,Fock1928,Darwin1930}. The magnetic field dependence is important to distinguish radiative Auger emission from phonon-related features.

\begin{figure*}
\includegraphics[width=1.8\columnwidth]{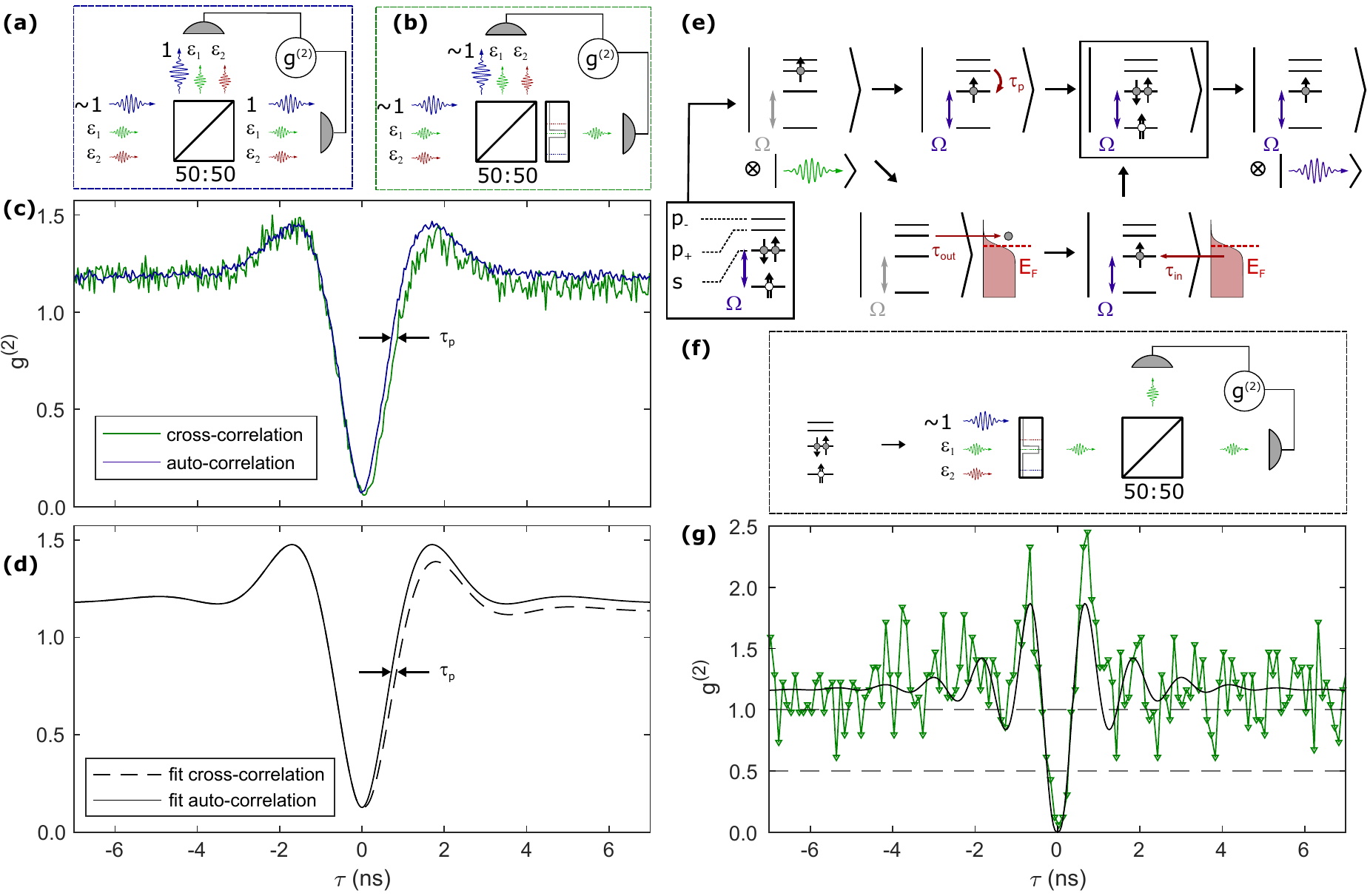}
\caption{\label{fig:crosscorr}{\bf Time-resolved correlation measurements. (a)} Schematic of the measurement to determine the auto-correlation of the resonance fluorescence from a quantum dot (QD). The signal is split by a 50:50 beamsplitter; photon arrival times are recorded on two single-photon detectors ($g^{(2)}$-measurement). {\bf (b)} Schematic of the cross-correlation measurement between resonance fluorescence and radiative Auger emission. The Auger emission is spectrally filtered to remove all resonant photons. {\bf (c)} Cross-correlation between the resonance fluorescence and the radiative Auger emission (green), {\color{black}measured on the InGaAs QD shown in Fig.\ \ref{fig:mechanism}}. An auto-correlation of the resonance fluorescence (blue) is shown for comparison. A time offset of $\tau_{p}\sim85\ \text{ps}$ between the auto-correlation and the cross-correlation is a measure of the relaxation time of a single electron from the $p$- to the $s$-shell. {\bf (d)} {\color{black}Fits to the auto- and cross-correlation measurements. (Parameters listed in \ref{tab:fitG2})} {\bf (e)} {\color{black}Model for the dynamics connected to the radiative Auger process.} After the radiative Auger excitation, the second electron occupies the $p$-shell of the QD. When the electron occupies the $p$- rather than the $s$-shell, the Coulomb interactions are different, tuning the $s$-to-$s$ transition out of resonance with the laser. The QD cannot be re-excited until the electron has relaxed to the $s$-shell. There are two relaxation channels: a direct relaxation to the $s$-shell on a time scale $\tau_p$; and ionization of the QD by tunneling from the $p$-shell to the Fermi-reservoir ($E_F$, Fermi energy) of the back gate ($\tau_{\text{out}}$) followed by slower tunneling from the Fermi reservoir to the $s$-shell ($\tau_{\text{in}}$). After relaxation, the QD is re-excited by the laser. {\bf (f)} Schematic setup for the auto-correlation measurement of the radiative Auger emission. The radiative Auger signal is split and sent to two single-photon detectors. {\bf (g)} Auto-correlation of the radiative Auger process involving the lower energy $p$-shell (green). The solid black line is a model where all parameters are identical to the simulation of the cross-correlation (see (d) and section \ref{sec:crosstheory}). Only the Rabi-frequency is higher compared to the cross-correlation measurement.}
\end{figure*}

The separation between resonance fluorescence and radiative Auger emission corresponds to the single-particle splittings. The radiative Auger lines, therefore, allow the single-particle spectrum of a quantum dot to be determined with high precision. At zero magnetic field ($B=0\ \text{T}$), there is a splitting between the two $p$-shell-related Auger lines, revealing an asymmetry of the QD. This asymmetry lifts the four-fold degeneracy of the $p$-shells into two doublets at zero magnetic field. For both types of QDs, we also observe radiative Auger emission at even lower energies (see Fig.\ \ref{fig:dispersion}(a,c)). These emission lines correspond to a radiative Auger process involving $d$-shells (Fig.\ \ref{fig:dispersion}(e)). At high magnetic fields, the upper $p$-shell ($p_-$) shows an anti-crossing with the lowest $d$-shell ($d_+$). For the GaAs QD, we even observe radiative Auger emission lines involving all three $d$-shells. For the InGaAs QD, the $d_+$-shell is only visible in the radiative Auger emission when it is coupled to the $p_-$-shell. For both types of QDs, we model the dispersion of the emission lines by the Fock-Darwin spectrum {\color{black}\cite{Fock1928,Darwin1930}} (details in \ref{sec:bfield}). The model assumes a harmonic confinement potential and matches well for the lower QD-shells (see Fig.\ \ref{fig:dispersion}(a,c)). Differences between model and data (e.g. for the $d$-shells) reveal the deviation from a harmonic confinement potential towards higher single-particle energy.

For a rotationally symmetric confinement potential, angular momentum is a good quantum number such that promotion of the Auger electron to the $d_0$-shell is possible, but promotion to the other $p$- and $d$-shells is forbidden. In practice, we find that the radiative Auger involving the $p$-shells is relatively strong and that the intensity of these processes is not strongly dependent on the magnetic field. Besides, the $p$-shells are not degenerate at zero magnetic field. These observations show that angular momentum is not a good quantum number. However, we do not observe Zeeman splittings in the radiative Auger lines, which shows that the processes are spin-conserving. Spin is a good quantum number; equivalently, spin-orbit interactions of the electron states are weak.

There are several additional red-shifted emission lines that are not related to electron shells or continuum states (see Fig.\ \ref{fig:dispersion}(a,b)): An emission red-shifted by $\sim36\ \text{meV}$ (labeled LO in Fig.\ \ref{fig:dispersion}(b,d)) corresponds to an optical recombination along with the creation of an LO-phonon (Fig.\ \ref{fig:dispersion}(f)). The magnetic field dispersion is weak and follows the QD $s$-shell -- no higher QD-shells are involved. At lower photon energies, even the LO-phonon replica of the radiative Auger emission is visible (labeled LO + $p_{\pm}$ in Fig.\ \ref{fig:dispersion}(b), schematic illustration in Fig.\ \ref{fig:dispersion}(g)). In this case, Auger carrier excitation into $p$-shells and LO-phonon creation occur simultaneously with the optical recombination. The identification of these lines is confirmed by the magnetic field dispersion which equals the dispersion of the radiative Auger emission (see Fig.\ \ref{fig:dispersion}(b)).

We turn to the dynamics of the radiative Auger process, that is, the dynamics of the electron left in an excited state after a radiative Auger process. Detecting a photon from a radiative Auger process projects the Auger electron into one of the excited electron states. The dynamics of this single electron can be investigated by determining the time of subsequent emission events. The experiment involves measuring the $g^{(2)}(\tau)$ correlation function with high precision in the delay $\tau$. We compare the auto-correlation of the resonance fluorescence (Fig.\ \ref{fig:crosscorr}(a)) to the cross-correlation between the radiative Auger emission and the resonance fluorescence (Fig.\ \ref{fig:crosscorr}(b)). This comparison provides immediate insight into the carrier relaxation mechanism following the radiative Auger process. The corresponding $g^{(2)}$-measurements are shown in Fig.\ \ref{fig:crosscorr}(c). 

The auto-correlation (blue curve) shows a very pronounced anti-bunching ($g^{(2)}<<1$) at zero time delay, proving the single-photon nature of the resonance fluorescence. The anti-bunching is surrounded by a bunching ($g^{(2)}>1$) at a non-zero time delay. This effect is caused by the onset of Rabi-oscillations under strong resonant driving. The cross-correlation (green curve) differs from the auto-correlation in two aspects: The $g^{(2)}(\tau)$ is a slightly asymmetric function of $\tau$ and has a time-offset towards positive $\tau$. We can explain these features {\color{black}(see fit in Fig.\ \ref{fig:crosscorr}(d)) with the mechanism shown in} Fig.\ \ref{fig:crosscorr}(e): After the emission of a radiative Auger photon, the second electron is located in a higher shell. Before re-excitation of the trion can take place, this electron has to relax down to the $s$-shell -- in contrast to the resonance fluorescence where re-excitation is immediately possible. By comparing auto- and cross-correlation, we determine the relaxation time for an isolated electron to be $\tau_{\text{p}}\simeq85\ \text{ps}$. The time-scale of the electron relaxation is comparable to numbers reported for weak nonresonant excitation \cite{Ohnesorge1996,Kurtze2009}. The relaxation is probably caused by a multi-phonon emission process \cite{Li1999}. We stress the advantage of the present method: the radiative Auger process leaves only a single electron in a higher shell. In contrast to nonresonant excitation, all other carriers have disappeared and the relaxation of the electron can be investigated independently of other relaxation mechanisms.

The asymmetry of the cross-correlation measurement can be explained by ionization of the QD following the radiative Auger emission. In a higher shell, the electron has an enhanced tunneling rate out of the QD \cite{Muller2013}. Following very fast relaxation down to the Fermi-energy, tunneling back into the $s$-shell of the QD takes about ten times longer, and the QD is ionized for a finite time. We estimate the corresponding tunneling times by modelling the auto- and cross-correlation measurements. {\color{black}The full model and the fit results are given in section \ref{sec:crosstheory}; the fits describe the experimental data well (see Fig.\ \ref{fig:crosscorr}(d))}.

Finally, we perform the first auto-correlation measurement of the radiative Auger emission. For this measurement, all the resonance fluorescence is filtered out (Fig.\ \ref{fig:crosscorr}(f)). To maximize the count rate of the weak radiative Auger emission, we use a higher Rabi-frequency compared to the cross-correlation measurement. The auto-correlation measurement is shown in Fig.\ \ref{fig:crosscorr}(g). At zero time delay, there is a clear anti-bunching in the $g^{(2)}$-measurement, which proves the single-photon nature of the emission connected to the radiative Auger process. At non-zero time-delay, the onset of Rabi-oscillations in the $s$-to-$s$ transition is visible as a photon bunching of the radiative Auger emission. Both features are well described by our model (section \ref{sec:crosstheory}).

The radiative Auger process takes place because the interactions between the carriers forming the trion change the eigenfunctions of the system (see section \ref{sec:theory}). In a single-particle basis, the initial state contains admixtures of Slater determinants{\color{black}\cite{Slater1929,Jackiw2018}} of higher single-particle shells. The optical recombination removes an electron-hole pair from the initial trion state, leading to a final state which is a superposition of single-electron single-particle states. Every state in that superposition consists of an electron in a particular shell along with a photon of a certain energy. Since the initial state is always the same, the energy separations between the different emission lines correspond to precise single-particle splittings. The ratio of radiative Auger emission and resonance fluorescence reflects the expansion of the trion state in single-particle states. Compared to the resonance fluorescence, the radiative Auger emission is weaker by about two to three orders of magnitude {\color{black}for both types of QDs}. {\color{black}It is slightly stronger for the larger GaAs QDs.} The trion wavefunctions are close, yet not equal to, single-particle states.

In conclusion, we experimentally studied negatively-charged trions in two different types of semiconductor QDs and observed a radiative Auger process in the optical recombination spectrum. We employ the radiative Auger process to determine the properties of a single electron in the QD -- the energy quantization and its relaxation and tunneling dynamics -- using the precise, sensitive and fast tools of quantum optics. The radiative Auger process only requires significant Coulomb interactions within the trion, a very general feature. Therefore, this process should also occur for the positively-charged trion and other quantum emitters in the solid-state.

\section{Acknowledgments}
\label{sec:acknowledgement}
We would like to thank Philipp Treutlein for fruitful discussions. MCL, CS, and RJW acknowledge financial support from NCCR QSIT and from SNF Project No.\ 200020\_156637. LZ received funding from the European Union Horizon 2020 Research and Innovation program under the Marie Sk\l{}odowska-Curie grant agreement No.\ 721394 (4PHOTON). AJ acknowledges support from the European Unions Horizon 2020 research and innovation program under the Marie Sk\l{}odowska-Curie grant agreement No.\ 840453 (HiFig). JR, AL, and ADW gratefully acknowledge financial support from the grants DFH/UFA CDFA05-06, DFG TRR160, DFG project 383065199, and BMBF Q.Link.X 16KIS0867. LM and PL gratefully acknowledges financial support from the Danish National Research Foundation (Center of Excellence Hy-Q, grant number DNRF139) and the European Research Council (ERC Advanced Grant SCALE).

\section{Author Contributions}
\label{sec:contrib}
MCL, CS, LZ, GNN, AJ performed the experiments. JR, ADW, and AL grew the samples. CS, MCL, LM fabricated the different samples. MCL, LZ, PL, AL designed the samples. MCL, CS, LZ, RJW analyzed the data. MCL developed the theory of the radiative Auger process. AJ, MCL, CS developed the theory for the time-resolved measurements. MCL, RJW, CS developed the theory for the magnetic field dispersion. MCL and RJW initiated the project and wrote the manuscript with input from all the authors.

\section{Methods}
\label{sec:methods}
The samples are grown by molecular beam epitaxy. Sample A contains InGaAs QDs embedded in a $p$-$i$-$n$-$i$-$n$-diode structure \cite{Lobl2017,Vora2015,Javadi2017,Grim2019}. Sample B contains GaAs QDs in AlGaAs, which are grown by GaAs-infilling of Al-droplet etched nano-holes \cite{Wang2007,Huo2013}. The photon out-coupling is enhanced by a distributed Bragg mirror below the QDs. For both samples, the QDs are placed between a $p$-doped top gate and an $n$-type doped back gate. The QDs are tunnel-coupled to the back gate. This configuration stabilizes the charge environment of the QDs and enables tuning the QD charge state by applying a voltage between top and back gate \cite{Patel2010,Kirsanske2017}. For the InGaAs QDs, the back gate has a distance of $40\ \text{nm}$ to the QDs, $30\ \text{nm}$ for the GaAs QDs. In a magnetic field, there is optical spin-pumping in the center of the trion charge plateau \cite{Kroutvar2004,Dreiser2008} (see section \ref{sec:spin}). Therefore, we perform all experiments at the plateau edges, where co-tunneling randomizes the electron spin \cite{Smith2005}.

All time-resolved measurements are performed by using superconducting single-photon detectors. The overall timing resolution for the $g^{(2)}$-measurements is $\text{IRF}\simeq35\ \text{ps}$ (full width at half maximum). Optical measurements are carried out at $4.2\ \text{K}$ in a helium bath cryostat. Resonant excitation of the QDs is performed with a narrow-bandwidth ($\sim1\ \text{MHz}$) tunable diode laser (Toptica DLpro), which is additionally filtered with a home-built grating setup in order to remove any background from the gain medium of the laser. Resonance fluorescence of individual QDs is measured by suppressing the reflected laser light with a cross-polarization technique.

\section{Modeling the Magnetic Field Dispersion}
\label{sec:bfield}
\begin{figure*}[t]
\includegraphics[width=1.8\columnwidth]{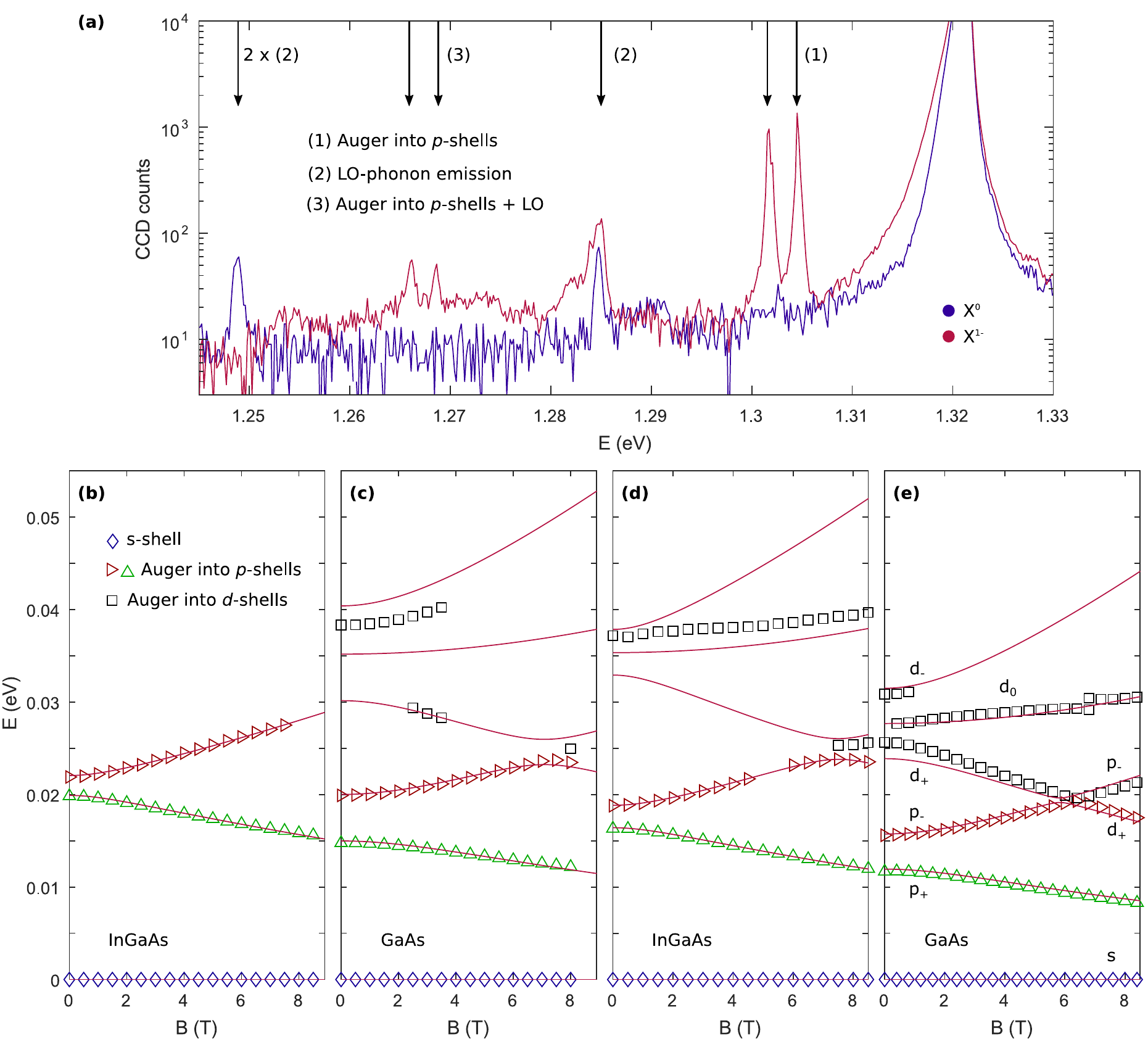}
\caption{\label{fig:mechanismFull}{\bf (a)} Emission spectrum of the InGaAs quantum dot shown in Fig.\ \ref{fig:dispersion}(a) in the main text. The $s$-to-$s$ transition of the QD is resonantly excited. The emission is shown for the neutral exciton (blue) and the singly charged trion (red). For better comparability, the neutral exciton ($X^0$) is shifted in energy such that its resonance fluorescence peak overlaps with the charged exciton ($X^{1-}$). In both cases, the resonance fluorescence (at $E\simeq1.321\ \text{eV}$) dominates. Emission at lower energies is caused by an energy transfer to either an LO-phonon or an additional carrier via the radiative Auger process. The radiative Auger process is only possible for the trion, not for the neutral exciton. {\bf(b)} Single-particle magnetic field dispersion for another InGaAs QD. To obtain the single-particle splittings, the $s$-shell energy is subtracted from the energies of the radiative Auger lines. {\bf(c)} Single-particle magnetic field dispersion for another GaAs QD embedded in AlGaAs. {\bf(d)} Single-particle splittings for the QD shown in Fig.\ \ref{fig:dispersion}(a) of the main text. {\bf(e)} Single-particle splittings for the QD shown in Fig.\ \ref{fig:dispersion}(c) of the main text.}
\end{figure*}

The radiative Auger emission appears on resonantly exciting the trion. Fig.\ \ref{fig:mechanismFull}(a) shows a comparison of the emission spectrum from a neutral exciton and a trion. The emission spectrum of the neutral exciton ($X^0$) only shows phonon-related features. A radiative Auger emission is impossible for the $X^0$ due to the absence of an Auger electron.

The final states after the optical recombination of the trion are single-particle states. Therefore, the separations between the different emission lines are precise single-particle splittings. Shown in Fig.\ \ref{fig:mechanismFull}(b,c) is the magnetic field dispersion of the extracted single-particle splittings for two different QDs. Fig.\ \ref{fig:mechanismFull}(d,e) shows the single-particle dispersion for the two QDs shown in the main text. At zero magnetic field, we measure an $s$-to-$p$-splitting of $17.7\ \text{meV}$ respectively $21.0\ \text{meV}$ on the InGaAs QDs; and $13.8\ \text{meV}$ respectively $17.6\ \text{meV}$ on the GaAs QDs. We can determine many further parameters of the single-particle spectrum by fitting the data to a model which assumes states of an asymmetric harmonic confinement potential. The red lines in Fig.\ \ref{fig:mechanismFull}(b-e) represent the model that is developed in this section. It reproduces the data very well.

For a symmetric, two-dimensional, and harmonic confinement potential, the magnetic field dependence of the single-particle states forms the Fock-Darwin spectrum \cite{Fock1928,Darwin1930}. The eigenergies $E_{n,L}$ depend on two quantum numbers, the radial quantum number, $n$, and the angular momentum quantum number, $L$ \cite{Kouwenhoven2001}. In this model, the two $p$-shells are degenerate at zero magnetic field. This is clearly not the case in our experiments. To describe the single-particle dispersions, we therefore assume an asymmetric harmonic confinement potential of the form $V(x,y)=\frac{1}{2}m_e^*\left(\omega_x^2x^2+\omega_y^2y^2\right)$. When the radial symmetry is broken, angular momentum is no longer a good quantum number, and the eigenenergies are $E_{n_x, n_y}=\hbar\omega_x\left(n_x+\frac{1}{2}\right)+\hbar\omega_y\left(n_y+\frac{1}{2}\right)$, with the two quantum numbers $n_x$ and $n_y$. The eigenenergies of such an asymmetric harmonic confinement as a function of the magnetic field are given in Ref.\ \onlinecite{Madhav1994}.

\begin{table*}
\begin{ruledtabular}
\begin{tabular}{ll} 
label & description\\\cline{1-2}
$\hbar$ & reduced Planck constant\\
$\mu_B$ & Bohr magneton\\
$\epsilon_0$, $\epsilon_r$ & permittivity of vacuum, relative permittivity\\
$g_e$, $g_h$ & electron and hole g-factor\\
$E_0$ & bandgap of the QD-material\\
$m_e^*$, $m_h^*$ & electron, hole effective mass\\
$\hbar\omega_c=\frac{\hbar eB}{m_e^*}$ & electron cyclotron energy\\
$\hbar\omega_x$, $\hbar\omega_y$ & confinement energies of the asymmetric harmonic oscillator\\
$\hbar\omega_0 \equiv\hbar\omega_{x}+\Delta_p\equiv\hbar\omega_{y}-\Delta_p$ & confinement energy of the symmetric harmonic oscillator\\
$n, L$ & quantum numbers for the symmetric harmonic oscillator\\
$n_x, n_y$ & quantum numbers for the asymmetric harmonic oscillator\\
$E_{n,L}$ & eigenenergies of the symmetric harmonic oscillator\\
$\Delta_{pd}$ & coupling between $p_-$- and the $d_+$-shell\\
$\Omega_{\text{R}}$ & Rabi frequency\\
$\Gamma_r=\tau_r^{-1}$ & radiative decay rate\\
$\Gamma_A=\tau_A^{-1}$ & radiative Auger decay rate\\
$\Gamma_p=\tau_p^{-1}$ & relaxation rate from $p$- to $s$-shell\\
$\Gamma_{\text{out}}=\tau_{\text{out}}^{-1}$ & tunnel rate out of the QD after a radiative Auger decay\\
$\Gamma_{\text{in}}=\tau_{\text{in}}^{-1}$ & tunnel rate into the ionized QD\\
$E_{\text{f}}^{p_{\pm}}, E_{\text{f}}^{d_{\pm}}$, and $E_{\text{f}}^{d_{0}}$ & final state energies after Auger excitation into $p$- and $d$-shells\\
$\Delta_{LO}, \Delta_{TO}$ & energies of longitudinal and transverse optical phonon
\end{tabular}
\caption{\label{tab:defs}List of definitions.}
\end{ruledtabular}
\end{table*}

The absolute energies of the emission lines correspond to the energy differences between the initial state ($E_{\text{trion}}$) and the final states ($E_{\text{f}}$). To fit the dispersions of these emission lines, we compute the energy of the initial trion state as the sum of its single-particle energies plus the corresponding Coulomb and exchange terms. For the Coulomb energy terms, we assume a symmetric confinement as the corresponding energy terms can be easily computed analytically \cite{Warburton1998,Cheng2003,Lobl2018}. Coupling terms admixing higher shells are not considered in this estimation \cite{Warburton1998,Cheng2003}.

At a magnetic field of $B\simeq8\ \text{T}$, the $p_-$- and the $d_+$-shells anticross. This is not a feature of the energy spectrum of an asymmetric harmonic oscillator. The anti-crossing is included by a phenomenological coupling $\Delta_{pd}$ between $p_-$- and $d_+$-shell. We speculate that the coupling between both shells arises due to the deviation from a harmonic confinement.

When part of the energy is transferred to an LO-phonon, the corresponding photon energy is given by, $E_{\text{trion}}-E_{\text{f}}^s-\Delta_{LO}$. This emission has the same weak magnetic field dependence as the resonance fluorescence ($s$-shell emission).

The results of fitting our model are shown in Fig.\ \ref{fig:mechanismFull} and Fig.\ \ref{fig:dispersion} of the main text. A list of definitions is given in Tab.\ \ref{tab:defs}, and the fit parameters are given in Tab.\ \ref{tab:fit}. For all measured QDs, the strong magnetic field dispersion of the radiative Auger emission lines is well reproduced.

In the case of the InGaAs QD shown in Fig.\ \ref{fig:dispersion}(a,b) (main text), we fit the energies of the $s$-shell emission and the radiative Auger emission into both $p$-shells simultaneously. The coupling term $\Delta_{pd}$ is included as a fit parameter. The exciton $g$-factor is measured independently by mapping out the charge plateau of the trion in a magnetic field. The fit reproduces the data very well and gives a good description of the radiative Auger excitation into some of the $d$-shells. The LO-phonon replica of the radiative Auger excitation into the $p$-shells is also excellently reproduced by the fit.

To fit the magnetic field dispersion of the InGaAs QD shown in Fig.\ \ref{fig:mechanismFull}(b), we also make a simultaneous fit to the energies of the $s$-shell emission and the radiative Auger emission into both $p$-shells. The coupling term $\Delta_{pd}$ is not included as there is no hint of an anticrossing with the $d_+$-shell.

For the GaAs QD shown in Fig.\ \ref{fig:dispersion}(c,d) of the main text, we again fit the energies of the $s$-shell emission and the radiative Auger emission into both $p$-shells simultaneously. The coupling term $\Delta_{pd}$ is included as a fit parameter. The exciton $g$-factor is measured independently and not fitted.

For the GaAs QD shown in Fig.\ \ref{fig:mechanismFull}(c), we also fit the energies of the $s$-shell emission and the radiative Auger emission into both $p$-shells simultaneously. The exciton $g$-factor is fixed to a value typical for GaAs QDs.

When observable, all phonon-related features are described using the fit results described above. A constant phonon energy is used as a single fit parameter. 

\begin{table*}
\begin{ruledtabular}
\begin{tabular}{lcccccccc} 
& $E_0$ (eV) & $m_e^*$ ($m_0$) & $g_h-g_e$ & $\hbar\omega_0$ (meV) & $\Delta_p$ (meV) & $\Delta_{pd}$ (meV) & $\Delta_{LO}$ (meV) & $\Delta_{TO}$ (meV)\\\cline{2-9}
InGaAs, Fig.\ \ref{fig:dispersion}(b) (main text) & 1.3214 & 0.0712 & 1.505 & 17.67 & 1.26 & 1.12 & 36.1 &--\\
GaAs, Fig.\ \ref{fig:dispersion}(d) (main text) & 1.5925 & 0.0757 & 1.135 & 13.84 & 1.90 & 0.25 & 36.3 & 33.5\\
InGaAs, Fig.\ \ref{fig:mechanismFull}(b) & 1.3152 & 0.0762 & 1.968 & 20.98 & 1.08 & -- &-- &--\\
GaAs, Fig.\ \ref{fig:mechanismFull}(c) & 1.5757 & 0.0737 & 1.1 & 17.59 & 2.61 & 1.37 & 36.5 &--
\end{tabular}
\caption{\label{tab:fit}Fit results for the magnetic field dispersion.}
\end{ruledtabular}
\end{table*}

\section{Radiative Auger Process: Theory}
\label{sec:theory}
To explain the radiative Auger process, we consider the interactions between the three particles forming the trion. We determine the multi-particle eigenstates, $\Psi$, for several carriers in the same QD by numerically solving the time-independent Schr\"odinger equation, $\hat{H}\Psi=E\cdot\Psi$, via exact diagonalization. The Hamiltonian, $\hat{H}$, of the system is:
\begin{equation}
\label{eq:H}
\hat{H}=\sum_{i=1}^N\left[\frac{-\hbar^2}{2m_i^*}\Delta_i + V\left(\vec{x}\right)\right] + \hat{C}.
\end{equation}
$\hat{C}$ is the Coulomb operator, which is given by:
\begin{equation}
\label{Eq:CoulombOp}
 \hat{C}=\frac{1}{4\pi\epsilon_0\epsilon_r}\sum_{i,\ j,\ i<j}^N\frac{c_i\cdot c_j}{\abs{r_i-r_j}}.
\end{equation}
The term $c_i=\pm e$ is the charge of a particle (electron or hole). As we are considering fermionic particles, the overall wavefunction is antisymmetric under particle exchange. Therefore, we consider $\hat{H}$ in a basis, $\{\Psi_{\textbf{n}}\}$, of antisymmetrized Slater determinants:
\begin{equation}
\Psi_{\textbf{n}}=\hat{\mathcal{A}}\prod_{i=1}^N\phi_{n_i}\left(x_i,\ \sigma_i\right).
\end{equation}
The Slater determinants are constructed from the single-particle solutions, $\phi_{n_i}\left(x_i,\ \sigma_i\right)$, of Eq.\ \ref{eq:H}. The index $\textbf{n}$ represents the quantum numbers required to describe all particles. The asymmetrization operator, $\hat{\mathcal{A}}$, constructs a Slater-determinant, which is asymmetric under the exchange of identical particles. To express $\hat{H}$ in the basis $\{\Psi_{\textbf{n}}\}$, the matrix elements $\bra{\Psi_{\textbf{n}}}{\hat{H}}\ket{\Psi_{\textbf{m}}}$ are computed. The Slater-Condon rules \cite{Slater1929,Condon1930} transform these multi-particle matrix elements into two-particle Coulomb matrix elements. The Slater-Condon rules for the two-particle Coulomb operator, $\hat{C}$, are:
\begin{align}
\label{Eq:SlCondon}
&\bra{\Psi_{\textbf{n}}}{\hat{C}}\ket{\Psi_{\textbf{n}}} = \frac{1}{2}\sum_{i,\ j,\ i\neq j}^N\left[V_{ijij} - V_{ijji}\right]\\
&\bra{\Psi_{\textbf{n}}}{\hat{C}}\ket{\Psi_{\textbf{n}\left(h,k\right)}} = \sum_{i=1}^N\left[V_{hiki}- V_{hiik}\right]\\
&\bra{\Psi_{\textbf{n}}}{\hat{C}}\ket{\Psi_{\textbf{n}\left(h,k,l,m\right)}} = V_{hlkm} - V_{hlmk}.
\end{align}
The index $\textbf{n}\left(h,k\right)$ indicates that this wavefunction is obtained from $\Psi_{\textbf{n}}$ by replacing the single-particle wavefunction $\phi_h$ of particle number $h$ by $\phi_k$. The index $\textbf{n}\left(h,k,l,m\right)$ means that two wavefunctions are changed correspondingly. The two-particle Coulomb matrix elements, $V_{hklm}$, are given by the following integral:
\begin{align}
\label{Eq:2pME}
&V_{hklm}=\bra{\phi_{h}\phi_{k}}{\hat{C}}\ket{\phi_{l}\phi_{m}}\nonumber\\
&\equiv\frac{e^2}{4\pi\epsilon_0\epsilon_r}\int\int \frac{\phi_h\left(\textbf{r}_1\right)^*\phi_k\left(\textbf{r}_2\right)^*\phi_l\left(\textbf{r}_2\right)\phi_m\left(\textbf{r}_1\right)}{\abs{\textbf{r}_1-\textbf{r}_2}}\ d\textbf{r}_1\ d\textbf{r}_2.
\end{align}
Depending on the order of the indices, these integrals include the direct Coulomb and the Coulomb exchange terms. For a symmetric harmonic confinement potential, analytic solutions for the Coulomb integrals can be found e.g. in Refs. \onlinecite{Warburton1998,Cheng2003}.

The eigenfunctions of Eq.\ \ref{eq:H} are obtained by diagonalizing $\hat{H}$ in the basis $\{\Psi_{\textbf{n}}\}$. The trion ground state has a small admixture of higher single-particle shells, which is the origin of the radiative Auger process. Upon optical recombination of one electron and a hole, the remaining electron of the trion is in a superposition including these higher shells. Detection of the frequency of the emitted photon projects the state of the remaining electron to the corresponding shell. For the trion, it is sufficient to carry out exact diagonalization for the initial state only since the final states are single-particle states.

In the dipole approximation, the emission spectrum can be computed with Fermi's golden rule \cite{Michler2003,Chithrani2005}:
\begin{equation}
I(\omega)\propto\sum_f\abs{\bra{\Psi^{(f)}}\hat{P}\ket{\Psi^{(i)}}}^2\cdot\delta(E_i-E_f-\hbar\omega)\cdot D(\omega),
\end{equation}
where $\Psi^{(i)}$ is the initial state, $\Psi^{(f)}$ are the possible final states, and $D(\omega)$ is the density of states for an emitted photon. $\hat{P}=\sum d_{ij}\hat{h}_{i,\sigma}\hat{e}_{j,-\sigma}$ adds up all dipole-matrix ($d_{ij}$) allowed electron-hole recombinations, where $i$, $j$ sum over orbital and $\sigma$ over spin degrees of freedom \cite{Michler2003,Chithrani2005}.

With the presented formalism, we estimate that the intensity of the radiative Auger transition from $s$- to the $d_0$-shell is about a hundred times weaker than the resonance fluorescence. However, this intensity is tendentially overestimated compared to the experimentally obtained values. The issue could be that the exact diagonalization only converges when taking into account very high single-particle shells. In reality, not all of these states exist due to close-by continuum states. {\color{black}Furthermore, the envelope wave approximation is a simplification compared to a fully atomistic treatment \cite{Cygorek2019}.} Finally, this approach assumes that angular momentum is a good quantum number, allowing radiative Auger with the $d_0$-shell but not with $p$-shells. In the experiment, radiative Auger with the $p$-shells is clearly observed, also in the limit of high magnetic field, suggesting that angular momentum is not a good quantum number.

\section{Cross-Correlation: Theory}
\label{sec:crosstheory}
\begin{figure*}[t]
\includegraphics[width=1.8\columnwidth]{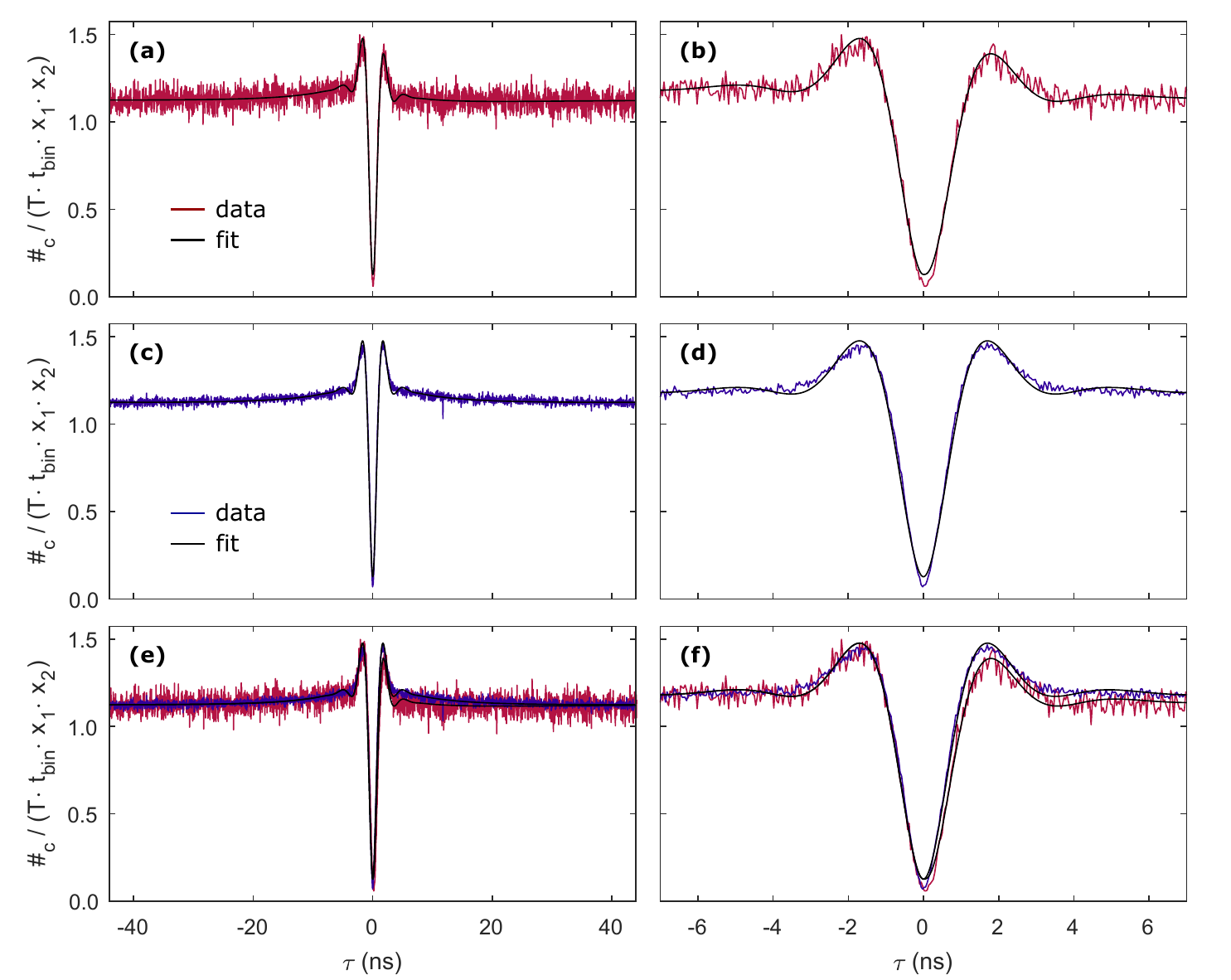}
\caption{\label{fig:g2Fit}Fits to the $g^{(2)}$-measurements shown in Fig.\ \ref{fig:crosscorr}(c) of the main text. Simultaneous fitting of Eqs.\ \ref{eq:autoCorr} and \ref{eq:crossCorr} to the corresponding auto- and cross-correlation data is performed. In the following sub-figures, the black lines correspond to the fit result. {\bf (a)} Red line: cross-correlation measurement between resonance fluorescence and the radiative Auger emission where the second electron is transferred into the $p_+$-shell of the quantum dot. {\bf (b)} Cross-correlation measurement from (a) on a shorter time-scale. {\bf (c)} Blue line: auto-correlation measurement of the resonance fluorescence. {\bf (d)} Auto-correlation measurement from (c) on a shorter time-scale. {\color{black}{\bf (e)} Comparison of the auto- and the cross-correlation measurement together with the corresponding fits. {\bf (f)} Comparison of the auto- and the cross-correlation measurement, plotted on a short time-scale.}}
\end{figure*}
\begin{figure}
\includegraphics[width=0.9\columnwidth]{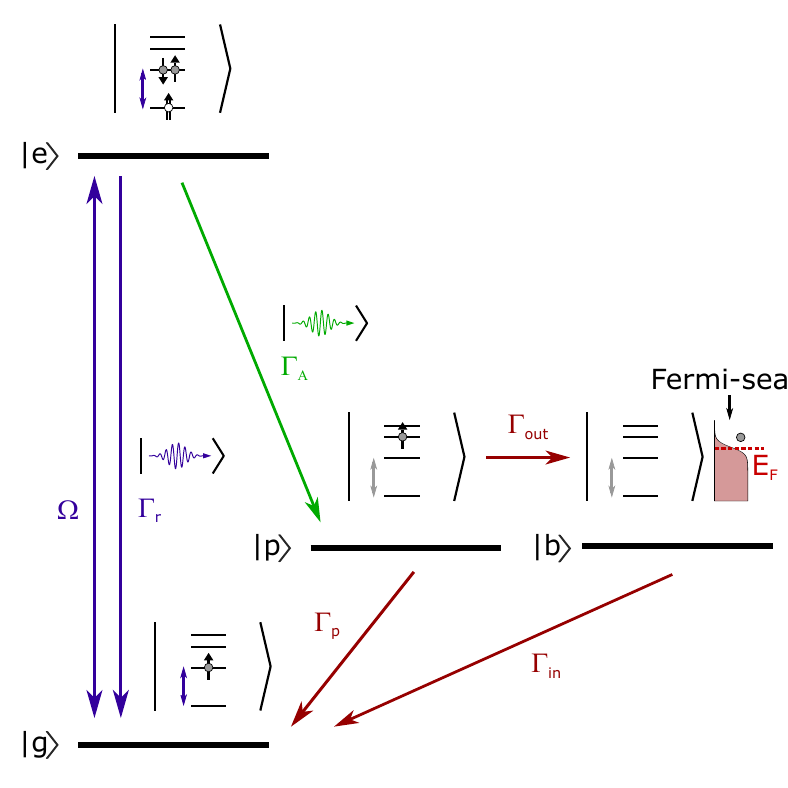}
\caption{\label{fig:g2Model}Model used for the simulation of the auto-correlation measurement of the resonance fluorescence together with the cross-correlation between the resonance fluorescence and the radiative Auger emission.}
\end{figure}
The $g^{(2)}$-measurements are modeled with the level scheme shown in Fig.\ \ref{fig:g2Model}. There are 4 different states which are taken into account for our simulation: the ground state, $\ket{g}$, with a single electron in the QD; the excited state, $\ket{e}$, a trion with two $s$-shell electrons; the state after a radiative Auger emission, $\ket{p}$, where a single electron occupies the $p$-shell of the QD; and the ionized QD-state, $\ket{b}$, where the electron has tunneled out of the QD. We simulate the system by assuming the Hamiltonian ($\hbar=1$):
\begin{equation}
\label{eq:HOmega}
\hat{H}=\frac{\Omega}{2}\left(\ket{g}\bra{e}+\ket{e}\bra{g}\right).
\end{equation}
All decay channels are modeled following the scheme shown in Fig.\ \ref{fig:g2Model}. The Lindblad operator is:
\begin{align}
\label{eq:collaps}
\hat{\mathcal{L}}&=\sqrt{\Gamma_r}\ket{g}\bra{e}+\sqrt{\Gamma_A}\ket{p}\bra{e}+\sqrt{\Gamma_p}\ket{g}\bra{p}\\
&+\sqrt{\Gamma_{\text{out}}}\ket{b}\bra{p}+\sqrt{\Gamma_{\text{in}}}\ket{g}\bra{b}.
\end{align}
We compute the steady-state density matrix, $\rho_s$, and obtain the auto- and cross-correlation by using the Quantum Toolbox in Python (QuTiP \cite{Johansson2013}). The operator for the resonant decay is $\hat{a}=\sqrt{\Gamma_r}\ket{g}\bra{e}$, and the operator for the radiative Auger decay is $\hat{a}_A=\sqrt{\Gamma_A}\ket{p}\bra{e}$. Auto- and cross-correlations are computed numerically by applying the quantum regression theorem. The auto-correlation of the resonance fluorescence is given by:
\begin{equation}
\label{eq:autoCorr}
g^{(2)}(\tau)=\frac{\langle \hat{a}^{\dagger}(t)\hat{a}^{\dagger}(t+\tau)\hat{a}(t+\tau)\hat{a}(t)\rangle}{\langle \hat{a}^{\dagger}(t)\hat{a}(t)\rangle^2}.
\end{equation}
The cross-correlation is given by:
\begin{equation}
\label{eq:crossCorr}
g^{(2)}(\tau)=\frac{\langle \hat{a}_A^{\dagger}(t)\hat{a}^{\dagger}(t+\tau)\hat{a}(t+\tau)\hat{a}_A(t)\rangle}{\langle \hat{a}^{\dagger}(t)\hat{a}(t)\rangle\langle \hat{a}_A^{\dagger}(t)\hat{a}_A(t)\rangle}.
\end{equation}
The auto-correlation of the radiative Auger emission is:
\begin{equation}
\label{eq:autoCorrAuger}
g^{(2)}(\tau)=\frac{\langle \hat{a}_A^{\dagger}(t)\hat{a}_A^{\dagger}(t+\tau)\hat{a}_A(t+\tau)\hat{a}_A(t)\rangle}{\langle \hat{a}_A^{\dagger}(t)\hat{a}_A(t)\rangle^2}.
\end{equation}
We multiply the result of this simulation by $1+c_1\cdot\exp\left(-\abs{\tau}/t_{bl}\right)$ to take into account a weak blinking on short time-scales \cite{Jahn2015}, which might be caused by electron spin pumping enabled by a weak nuclear magnetic field \cite{Hansom2014}. Additionally, the model function is multiplied with a global prefactor $c_0$, which takes into account a weak blinking on a time-scale of $\sim0.1\ \text{ms}$, probably caused by charge noise. For the resonance fluorescence, a small fraction $c_l$ of reflected laser in the resonant emission is taken into account via $g^{(2)} \rightarrow g^{(2)}\cdot(1-c_l)+c_l$. We perform a simultaneous fit of this model to the auto-correlation of the resonance fluorescence and the cross-correlation between the resonance fluorescence and the radiative Auger emission. The result of this fit is shown in Fig.\ \ref{fig:g2Fit}. The obtained fit parameters are stated in Tab.\ \ref{tab:fitG2}. These parameters also give a good fit to the auto-correlation of the radiative Auger emission, which is shown in Fig.\ \ref{fig:crosscorr}(g) of the main text. All fit parameters are kept the same, and only the Rabi-frequency is increased ($\Omega_{\text{R}}=5.4\ \text{GHz}$), taking into account that the auto-correlation of the radiative Auger emission has been measured at higher power.

\begin{table*}
\begin{ruledtabular}
\begin{tabular}{cccccccccc} 
$\Omega$ (GHz) & $\Gamma_r$ (GHz) & $\Gamma_A$ (GHz) & $\Gamma_p$ (GHz) & $\Gamma_{\text{out}}$ (GHz) & $\Gamma_{\text{in}}$ (GHz) & $t_{bl}$ (ns) & $c_0$ & $c_1$ & $c_l$\\\cline{1-10}
1.85& 1.22 & 0.001 &11.7 & 0.82 & 0.07 & 7.2 & 1.143&0.153&0.126
\end{tabular}
\caption{\label{tab:fitG2}Parameters obtained from simultaneously fitting the auto- and cross-correlation measurements shown in Fig.\ \ref{fig:g2Fit}. The radiative decay rate, $\Gamma_r$, is obtained from a separate lifetime measurement and is not included in the fit. $\Gamma_A$ is estimated from the intensity ratio between radiative Auger emission and resonance fluorescence and is also not included in the fit.}
\end{ruledtabular}
\end{table*}

\section{Evaluation of Correlation Measurements}
All $g^{(2)}$-measurements are performed in a time-tagged, time-resolved mode. The arrival times of all photons are recorded over the full integration time, $T$, on two single-photon detectors. Any analysis is carried out post-measurement. We compute the cross-correlation ($g^{(2)}$) between both signals by counting the coincidence events between the two detectors as a function of a time delay, $\tau$, between the signals.

Let $x_{1}$, $x_{2}$ be the count rates on detectors 1 and 2, respectively. We divide the full integration time into time-intervals of length, $t_{bin}$. The value for $t_{bin}$ is chosen to be small enough such that the probability of a photon in the corresponding time-interval is very small: $t_{bin}\cdot x_{1/2}\ll1$. For each detector, we determine the number of detection events in every small time interval. This number is either 0 for no photon or 1 for one photon since the probability of having more than one photon in an interval is negligibly small (for $t_{bin}\cdot x_{1/2}\ll1$). When there is one detection event on detector 1 in an interval at time $t$ and another detection event on detector 2 in an interval at time $t+\tau$, we call it a coincidence event for time delay $\tau$. For different time delays, we count the number of coincidence events, $\#_c$, over the full integration time. The cross-correlation between both detectors is obtained by dividing $\#_c(\tau)$ by its expectation value for the case of two uncorrelated detection channels: $\langle\#_c\rangle=T\cdot t_{bin}\cdot x_1 \cdot x_2$. This expression for $\langle\#_c\rangle$ is obtained by the following consideration: the probability of finding a detection event in a certain time interval is $t_{bin}\cdot x_{1}$ and $t_{bin}\cdot x_{2}$. If both detection channels are uncorrelated, the probability of finding a detection event for the first detector at time $t$ and a detection event for the second detector in the time-interval at $t+\tau$ is $p_c=t_{bin}^2\cdot x_{1}\cdot x_{2}$. For $T\gg\tau$, the probability density distribution of $\#_c$ is thus a binomial distribution:
\begin{equation}
\label{eq:countDistrib}
P(\#_c)={T/t_{bin}\choose \#_c} \cdot\left(1-p_c\right)^{T/t_{bin}-\#_c}\cdot p_c^{\#_c}
\end{equation}
The expectation value of this distribution is the corresponding normalization factor: $\langle\#_c\rangle=T\cdot t_{bin}\cdot x_1 \cdot x_2$.

\section{Power Dependent Excitation}
\label{sec:power}
\begin{figure}[b]
\includegraphics[width=0.9\columnwidth]{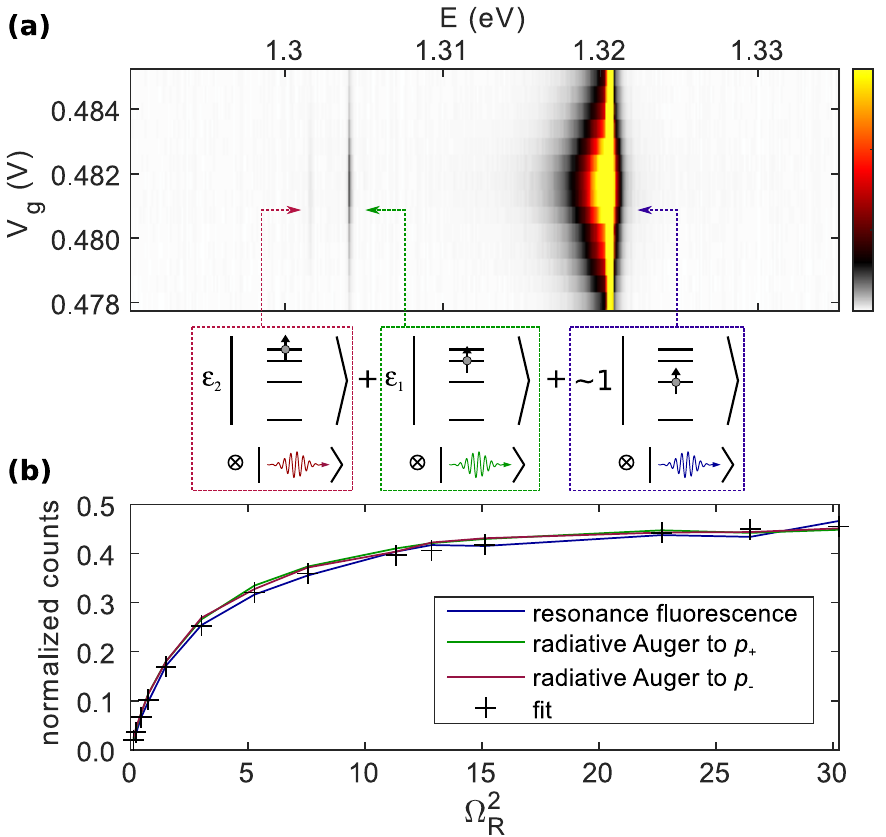}
\caption{\label{fig:pwr}{\bf (a)} Resonance fluorescence and radiative Auger emission. The excitation laser is fixed ($E\simeq1.321\ \text{meV}$), and the QD is swept through the resonance by tuning the gate voltage, $V_g$. {\bf (b)} Dependence of resonance fluorescence and radiative Auger emission on the power of the resonant laser. For the power dependence, the laser is kept on resonance with the trion ($X^{1-}$). When normalized, the resonance fluorescence and the radiative Auger emission intensity depend equally on the excitation power. Both are proportional to the upper state occupation of a resonantly driven two-level system (Eq.\ \ref{Eq:p22}).}
\end{figure}
We measure the intensity of the radiative Auger emission as a function of resonant excitation power and laser detuning. This measurement is shown in Fig.\ \ref{fig:pwr}. In a first measurement, we keep the narrow-band laser at a fixed frequency and sweep the detuning between trion transition and laser by applying a gate voltage, $V_g$. The gate voltage shifts the trion energy via the quantum-confined Stark effect. The intensity and the energy of the emission are recorded on a spectrometer. This measurement is shown in Fig.\ \ref{fig:pwr}(a). When laser and trion energy are on resonance, there is a bright emission at $\sim1.321\ \text{eV}$, the resonance fluorescence. This emission is spectrally asymmetric due to the LA-phonon sideband around the resonant peak. At lower energy, $\sim18\ \text{meV}$ below the resonance fluorescence, there is the emission corresponding to the radiative Auger excitation into the $p$-shells. This emission is strongest when also the resonance fluorescence is at its maximum, indicating that the intensity of the radiative Auger emission is proportional to the excited state population of the QD. Our model of the radiative Auger process implies this proportionality since the process only takes place in the excited state (trion) of the QD.

To investigate this dependence further, we keep the laser on resonance with the trion and measure the emission intensities as a function of power. This measurement is shown in Fig.\ \ref{fig:pwr}(b). The power dependence of the resonance fluorescence and the radiative Auger emission follows the power saturation curve of a two-level system very well. This result also confirms that the radiative Auger process is entirely related to the trion. Its rate is proportional to the trion occupation, $\rho_{22}$, under resonant excitation \cite{Loudon2000}:
\begin{equation}
\label{Eq:p22}
\rho_{22}=\frac{1}{2}\frac{\Omega_{\text{R}}^2}{2\Gamma_r^2+\Omega_{\text{R}}^2}.
\end{equation}

We expect that the ratio of the radiative Auger and the resonance fluorescence intensities roughly reflects the ratio $\Gamma_A/\Gamma_r$. This way, we estimate the value for $\Gamma_A$ to be on the order of $\sim1\ \text{MHz}$.

\begin{figure}[t]
\includegraphics[width=0.9\columnwidth]{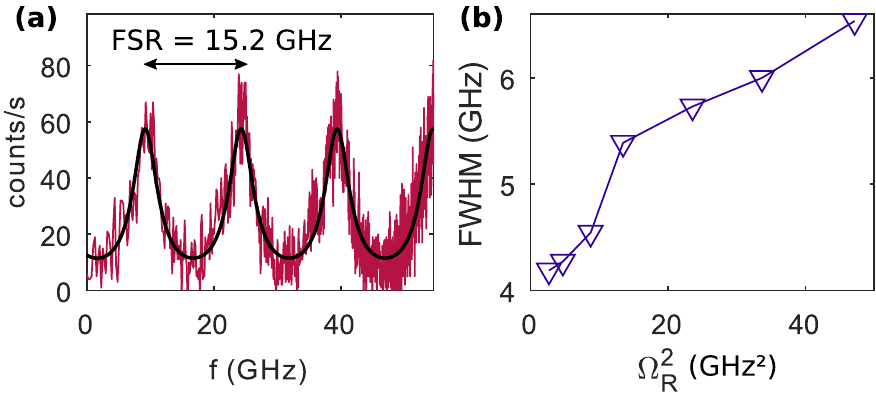}
\caption{\label{fig:FP}{\bf (a)} Radiative Auger emission at $\Omega_R=0.73\ \text{GHz}$ transmitted trough a $0.41\ \text{GHz}$ Fabry-Perot cavity. {\bf (b)} Linewidth of the radiative Auger emission as a function of the resonant Rabi frequency.}
\end{figure}

Finally, we measure the linewidth of the radiative Auger emission. We pass the emission through a Fabry-Perot cavity ($15.2\ \text{GHz}$ free spectral range, $0.41\ \text{GHz}$ linewidth) and sweep the cavity length. The result of this measurement on the $p_+$-emission is shown in Fig.\ \ref{fig:FP}(a). We determine the linewidth of the radiative Auger emission by fitting a multi-Lorentzian which is convoluted with the cavity linewidth. At low power, we measure a minimum linewidth of $4.19\ \text{GHz}$. For comparison, the lifetime limited linewidth is estimated by the decay rate of the $p_+$-state after the radiative Auger process: $\frac{\Gamma_r}{2\pi}=1.99\ \text{GHz}$. We repeat the linewidth measurement for different excitation Rabi frequencies. This measurement is shown in Fig.\ \ref{fig:FP}(b) and shows a linear increase of the linewidth as a function of the excitation power. The reason for the additional contribution to the linewidth and its linear broadening with the excitation power requires further investigations.

\begin{figure*}[b]
\includegraphics[width=1.8\columnwidth]{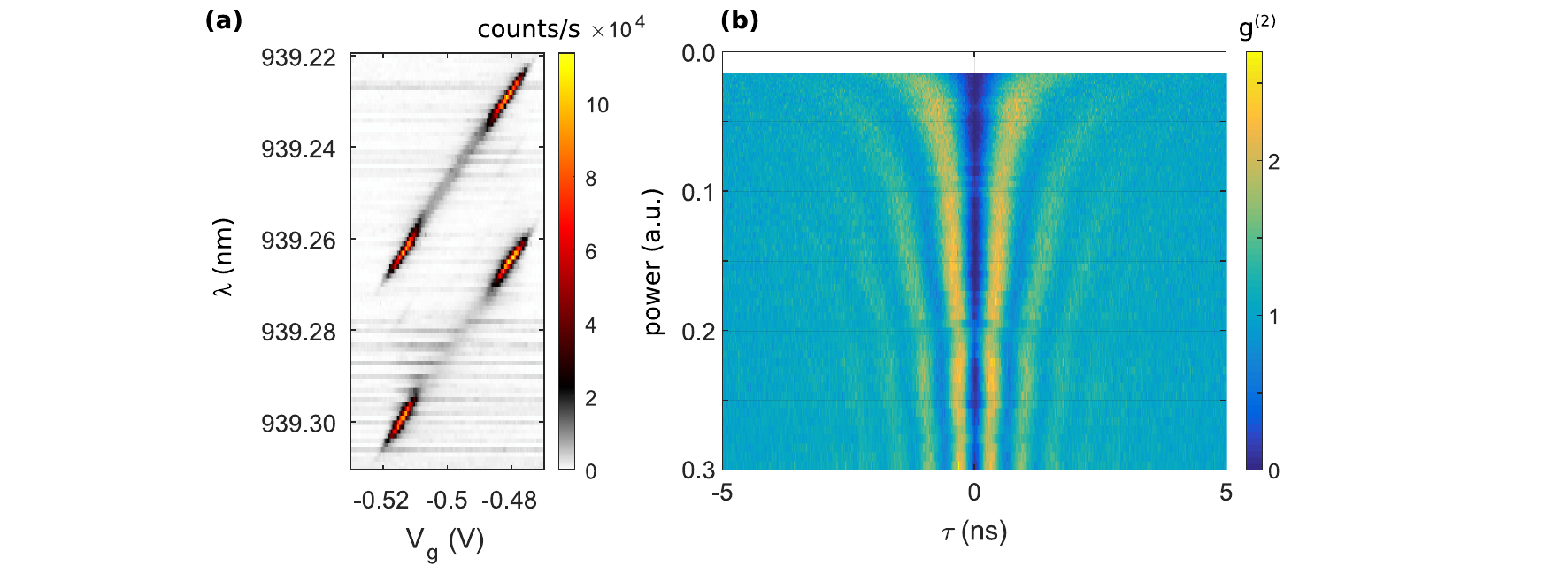}
\caption{\label{fig:plateau}{\bf (a)} The charge plateau of the resonantly excited trion at a magnetic field of $0.6$ T. This measurement is carried out on the InGaAs quantum dot (QD) shown in Fig.\ \ref{fig:mechanism}(b) of the main text. At the edges of the charge plateau, a strong resonance fluorescence is detected. In the plateau center, the resonance fluorescence intensity is strongly reduced due to electron spin pumping. All radiative Auger measurements are performed at the plateau edges. {\bf (b)} Power dependent $g^{(2)}$-measurement on the negative trion of the same QD.}
\end{figure*}

\section{Spin pumping and Rabi oscillations}
\label{sec:spin}
Fig.\ \ref{fig:plateau}(a) shows a measurement of the resonance fluorescence of the negative trion as a function of the gate voltage and the laser wavelength. This measurement is performed on the quantum dot which is presented in Fig.\ \ref{fig:mechanism}(b) of the main text. The trion is stable in the gate voltage range between $V_g=-0.52\ $V and $V_g=-0.48\ $V. This charge plateau splits into two due to the electron spin Zeeman energy. We perform the measurements of the radiative Auger emission on one Zeeman branch. No Zeeman splitting is observed in the emission spectrum, which shows that the radiative Auger process is spin-conserving. In the center of the charge plateau, the resonance fluorescence disappears due to optical spin pumping. At the edges of the charge plateau, the resonance fluorescence is strong due to spin co-tunneling with the back gate \cite{Smith2005,Dreiser2008}. For this reason, we perform all measurements in the co-tunneling regime.

Fig.\ \ref{fig:plateau}(b) shows resonantly driven Rabi-oscillations as a function of the excitation power. The measurement is performed on the trion state of the same QD. These coherent oscillations in the auto-correlation ($g^{(2)}$) measurement show that the QD can be approximately described by a two-level system \cite{Flagg2009}. However, radiative Auger is a fundamental process that limits this two-level approximation in the case of a trion.

\bibliography{Auger2019_v6_arxiv.bbl}

\end{document}